\documentclass{PoS}

\newcommand{\rgl}{\rangle}
\newcommand{\lgl}{\langle}
\usepackage{color}
\usepackage{xcolor}

\newcommand{\w}{\textcolor{white}}
\newcommand{\hdistance}{\hspace{3.5mm}}
\newcommand{\vdistance}{\vspace{2mm}}

\title{Critical slowing down and the gradient flow coupling in the Schr\"odinger functional}

\ShortTitle{Critical slowing down and the gradient flow coupling in the Schr\"odinger functional}

\author{\hfill\parbox{3cm}{\small\it%
DESY 13-191   \\
HU-EP-13/54   \\
SFB/CPP-13-77 \\
}}

\author{Patrick Fritzsch\\
        Humboldt-Universit\"at, Institut f\"ur Physik, Newtonstr. 15, 12489 Berlin, Germany \\
        E-mail: \email{fritzsch@physik.hu-berlin.de}}
        
\author{Alberto Ramos\\
        NIC, DESY, Platanenallee 6, 15738 Zeuthen, Germany \\
        E-mail: \email{alberto.ramos@desy.de}}

\author{\speaker{Felix Stollenwerk}%
\\
        Humboldt-Universit\"at, Institut f\"ur Physik, Newtonstr. 15, 12489 Berlin, Germany \\
        E-mail: \email{felix.stollenwerk@physik.hu-berlin.de}}
        
\abstract{
We study the sensitivity of the gradient flow coupling to sectors of different topological charge and its implications in practical situations.
Furthermore, we investigate an alternative definition of the running coupling that is expected to be less sensitive to the problems of the
HMC algorithm to efficiently sample all topological sectors.
}

\FullConference{31st International Symposium on Lattice Field Theory LATTICE 2013\\
		 July 29 -- August 3, 2013\\
		 Mainz, Germany}

\begin{document}

\section{Introduction}
The gradient (or Wilson) flow in Yang-Mills theory is
defined by the non-linear equation \cite{luescher10_flow}
\begin{equation}
 \frac{dB_\mu(x,t)}{dt} = D_\nu G_{\nu\mu}(x,t) \ , \qquad  B_\mu(x,0) = A_\mu(x) \ ,
 \label{eq:floweq}
\end{equation}
where $t$ is the flow time and $G_{\mu\nu} = \partial_\mu B_\nu - \partial_\nu B_\mu + [B_\mu, B_\nu]$. 
Due to 
\begin{equation}
 D_\nu G_{\nu\mu} \sim -\frac{\delta S_{\rm YM}[B]}{\delta B_\mu} \ ,
\end{equation}
the flow effectively constitutes {a smoothing of the gauge field over a range $\sqrt{8t}$} and
drives it towards a local minimum of the action. 
One remarkable feature of this procedure is the fact that 
the gauge field $B_\mu$ at $t > 0$ is renormalized, rendering
       expectation values of local, gauge-invariant quantities like the energy density
       \begin{equation}
        \lgl E(t) \rgl = \frac14 \lgl G_{\mu\nu}^a (t) G_{\mu\nu}^a (t) \rgl 
        \label{eq:energydensity}
       \end{equation}
       finite \cite{luescher10_flow}.
This can be used to define a
non-perturbative gradient flow (GF) coupling in terms of 
the energy density at positive flow time \cite{luescher10_flow}
at a renormalization scale 
{$\mu = 1/\sqrt{8t}$}.
In order to study the running coupling in asymptotically free theories such as QCD,
finite size scaling is applied, 
i.e., the renormalization scale runs with the size $L$ of the box,
\begin{equation}
 {\mu = \frac{1}{\sqrt{8t}}} = \frac{1}{cL} \ ,
 \label{eq:fss}
\end{equation}
{where $c$ represents the fraction of the box over which the gauge field is smoothed}.
This approach was implemented in a periodic box \cite{Fodor:2012td} and in the Schr\"odinger functional (SF) \cite{fritzsch13_flowNf2},
where the GF coupling is defined as
\begin{equation}
   \bar g^2_{\rm GF} (L) := \mathcal N^{-1} \cdot t^2 \lgl E(t,x_0) \rgl \big|_{t=c^2L^2/8}^{x_0 = T/2} \ .  
   \label{eq:gGF}
\end{equation}
The normalization factor $\mathcal N^{-1}$ in (\ref{eq:gGF}) ensures
$\bar g^2_{\rm GF} = g_0^2 + \mathcal O(g_0^4)$.
Note that as the SF breaks translational invariance in time direction, the energy density becomes explicitly 
dependent\footnote{Boundary fields and fermionic phase angle of the SF need to be specified as well.}
on $x_0$.

In \cite{fritzsch13_flowNf2}, the GF coupling has been investigated on the lattice 
in view of numerical costs and
cutoff effects for ensembles of $N_f=2$ simulations at a line of constant physics, defined by a constant
value of the SF coupling, which corresponds to $L \sim 0.4~{\rm fm}$.
Smoothing fractions in the range $c\in[0.3,0.5]$ turn out to be convenient, as their use leads to 
high statistical precision at affordable cost and
modest cutoff effects.
However, at the considered physical volume, the path integral is largely dominated 
by the trivial topological sector ($Q=0$) and contributions from other
sectors can be 
considered negligible. 

In contrast, at larger volumes like $L \sim 0.8~{\rm fm}$, sectors of non-vanishing topological charge 
are expected to contribute,
which brings up the question about the well-known problems of topology freezing and critical slowing down~\cite{Schaefer:2010hu}.
The aim of the present work is to investigate whether and how the determination of the gradient flow coupling
is affected by these phenomena.

\section{Numerical simulations}

In order to be able to produce large statistics, we perform simulations in pure SU(3) Yang-Mills theory with the Wilson gauge action. 
We choose SF boundary conditions with zero boundary gauge fields,
and fix the physical volume in terms of the Sommer scale:
\begin{equation}
 L = r_0 / 0.563 \sim 0.8~{\rm fm} \ .
 \label{eq:L}
\end{equation}
For lattice sizes of $L/a = 8,12,16,20,24$, we simulate along the line of constant physics defined by
(\ref{eq:L}).
The corresponding bare couplings ($\beta = 6/g_0^2$) are determined using \cite{necco01_latticescaleNf0}
\begin{equation}
  \log \left( a/r_0 \right) = -1.6804 - 1.7331(\beta-6) +0.7849(\beta-6)^2 - 0.4428(\beta-6)^3
  \label{eq:scalesetting}
\end{equation}
and can be found along with other parameters of the simulation in Tab.~\ref{tab:numerical_simulations}.
We use the HMC of the openQCD package \cite{luescher_openQCD}
and the 2-loop value of $c_t$ for $\mathcal O(a)$ improvement \cite{Bode:1999sm}.
Each produced configuration is evolved by integration of the flow equation (\ref{eq:floweq}),
including flow times $t$ corresponding to $c=0.3, 0.5$. 
Afterwards, on the smoothed configurations the topological charge, 
\begin{equation}
 Q(t) = \frac{1}{16\pi^2} \sum_{x} G_{\mu\nu}(x,t) \tilde G_{\mu\nu}(x,t)
\ ,
\end{equation}
and the GF coupling are measured
using the clover discretisation for the field strength.

\begin{table}
\begin{minipage}{0.48\textwidth}
\centering
\small
\begin{tabular}{c|c|c|c}
$L/a$          & $\beta$ & $N_{\rm ms}$ & MDUs \\ \hline
  8            & 5.9032 & 80000 & 6 \\
{12}       & 6.1410 & 80000 & 6 \\
{16}    & 6.3413 & 40000 & 6 \\
{20} & 6.5119 & 15000 & 12 \\
{24}      & 6.6552 & ~7000 & 12
\end{tabular}
\caption{Parameters of the numerical simulations. $N_{\rm ms}$ is the amount of measurements 
         and the last column shows the number of molecular dynamic units between two consecutive measurements. 
         }
\label{tab:numerical_simulations}
\end{minipage}
\begin{minipage}{0.04\textwidth}
 \w{a}
\end{minipage}
\begin{minipage}{0.48\textwidth}
\centering
\small
\begin{tabular}{c|c|c}
$L/a$ & $c=0.3$          & $c=0.5$          \\ \cline{1-3}
  8   & 98.66(\w{0}8)    & 98.94(\w{0}8)    \\
{12}  & {98.19(20)}      & {98.45(19)}      \\
{16}  & {98.46(62)}      & {98.56(61)}      \\
{20}  & {*99.91(\w{0}3)} & {*99.96(\w{0}2)} \\
{24}  & {*99.52(36)}     & {*99.54(36)}
\end{tabular}
\caption{Percent of configurations with topological charge $Q \leq 0.5$.
         The values denoted with an asterisk are biased.
         \newline 
}
\label{tab:Q_distribution}
\end{minipage}
\end{table}

\section{Results}

\subsection{Distribution of the topological charge \label{ss:Q}}

Histories of the topological charge are shown in
Fig.~\ref{fig:Q_distribution} for $c=0.3$.
\newcommand{\scaleA}{0.37}
\begin{figure}
\centering
\includegraphics[scale=\scaleA]{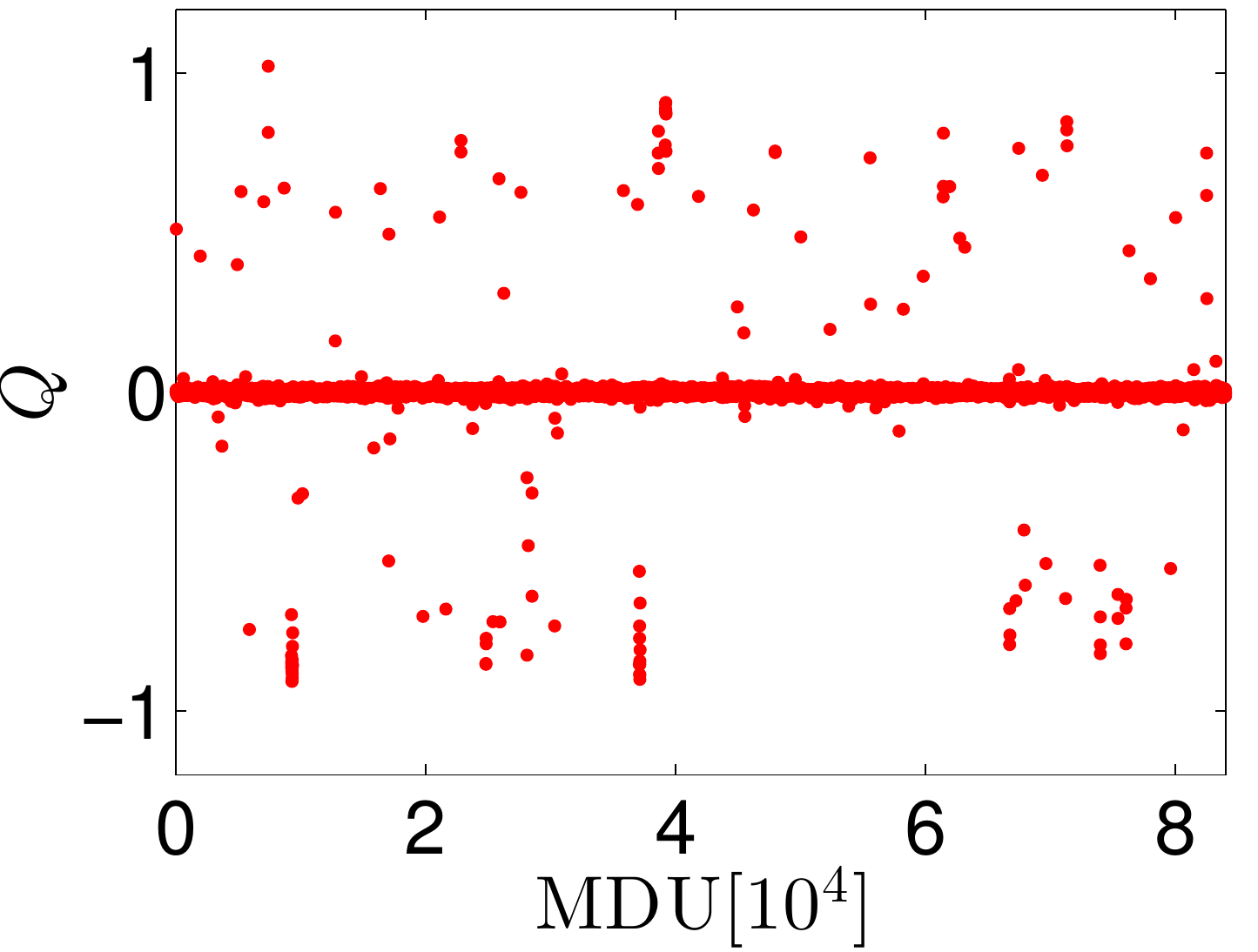}
\hdistance 
\includegraphics[scale=\scaleA]{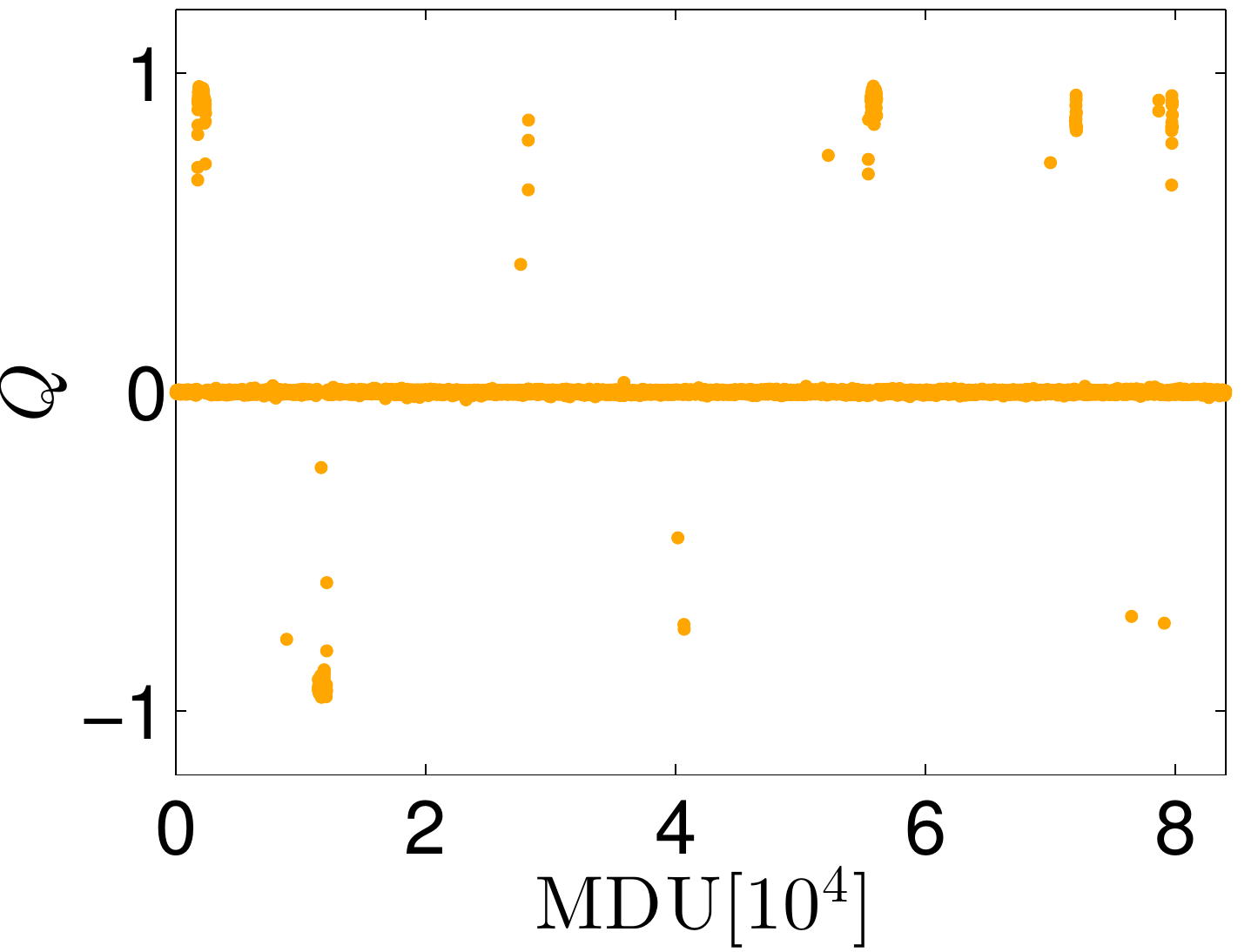} \\
\vdistance \vspace{1mm}
\includegraphics[scale=\scaleA]{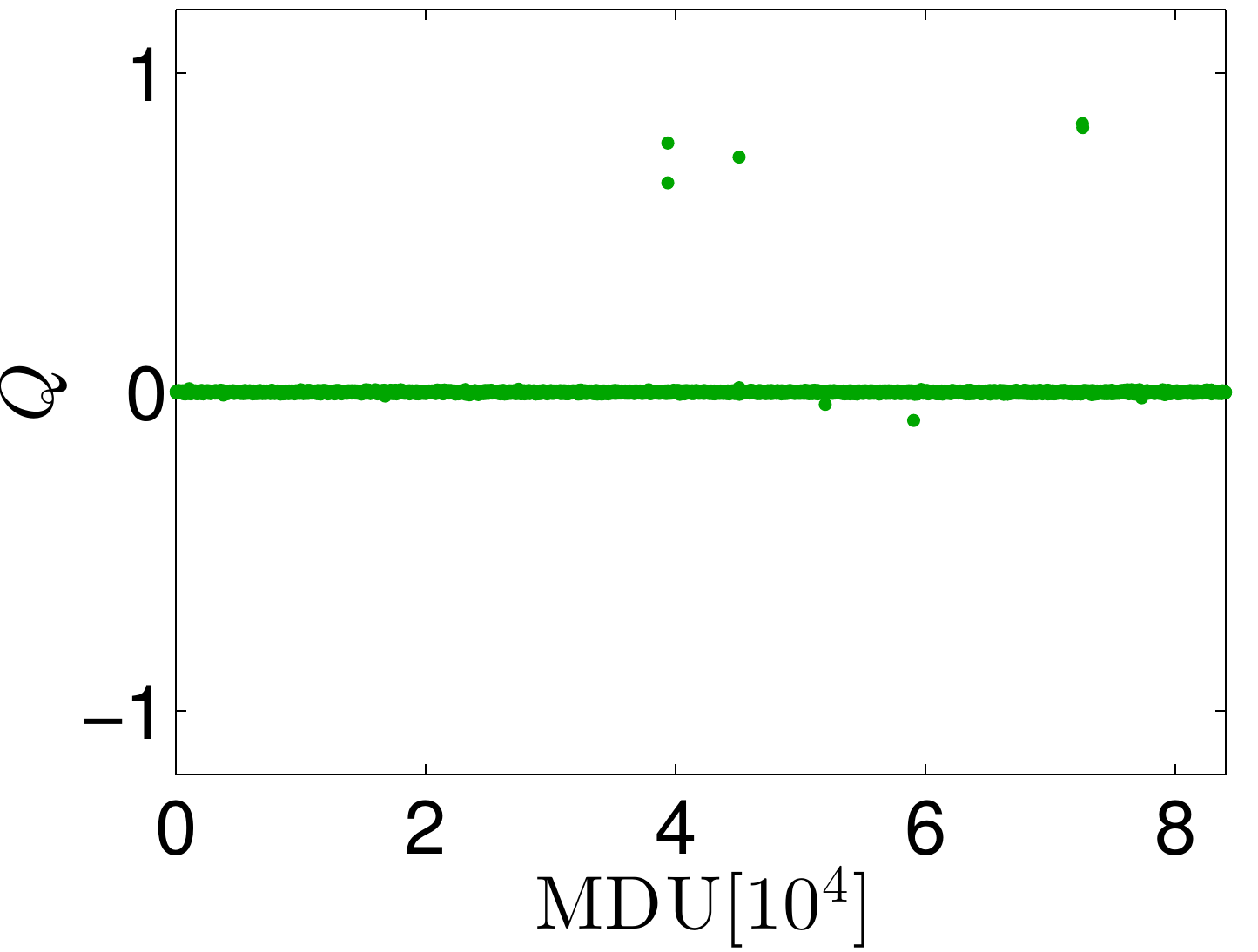}
\hdistance
\includegraphics[scale=\scaleA]{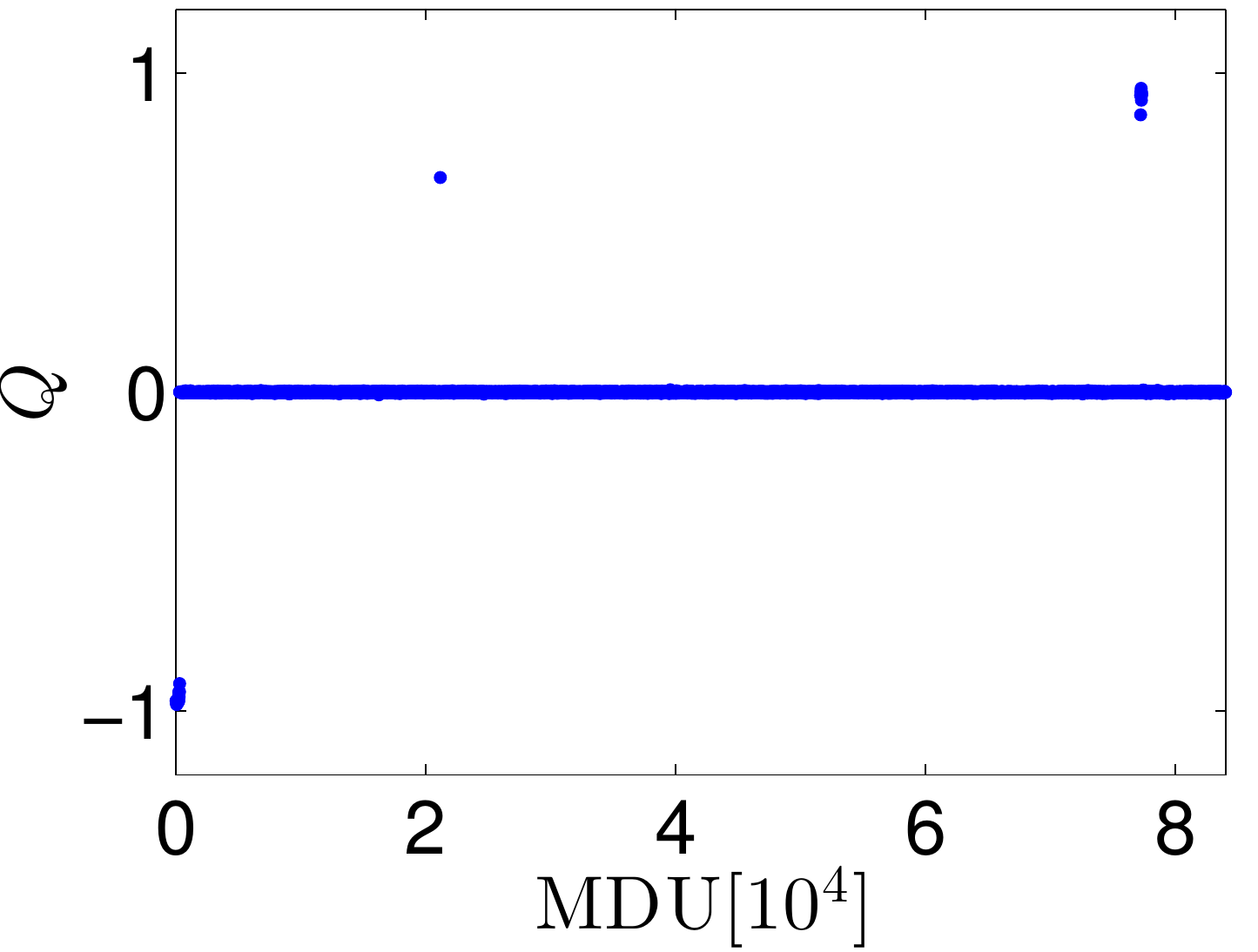}
\caption{Histories (in excerpts) of $Q$, for $c=0.3$. The plots in red, orange, green and blue correspond to $L/a = 12,16,20,24$, respectively.
}
\label{fig:Q_distribution}
\end{figure} 
For lattices up to $L/a=16$, one observes that non-trivial configurations appear to cluster more and more 
as the lattice gets finer. 
This goes together with the increasing integrated autocorrelation time 
displayed in Fig.~\ref{fig:autocorrelation_Q}. 
\newcommand{\scaleB}{0.37}
\begin{figure}
\centering
\includegraphics[scale=\scaleB]{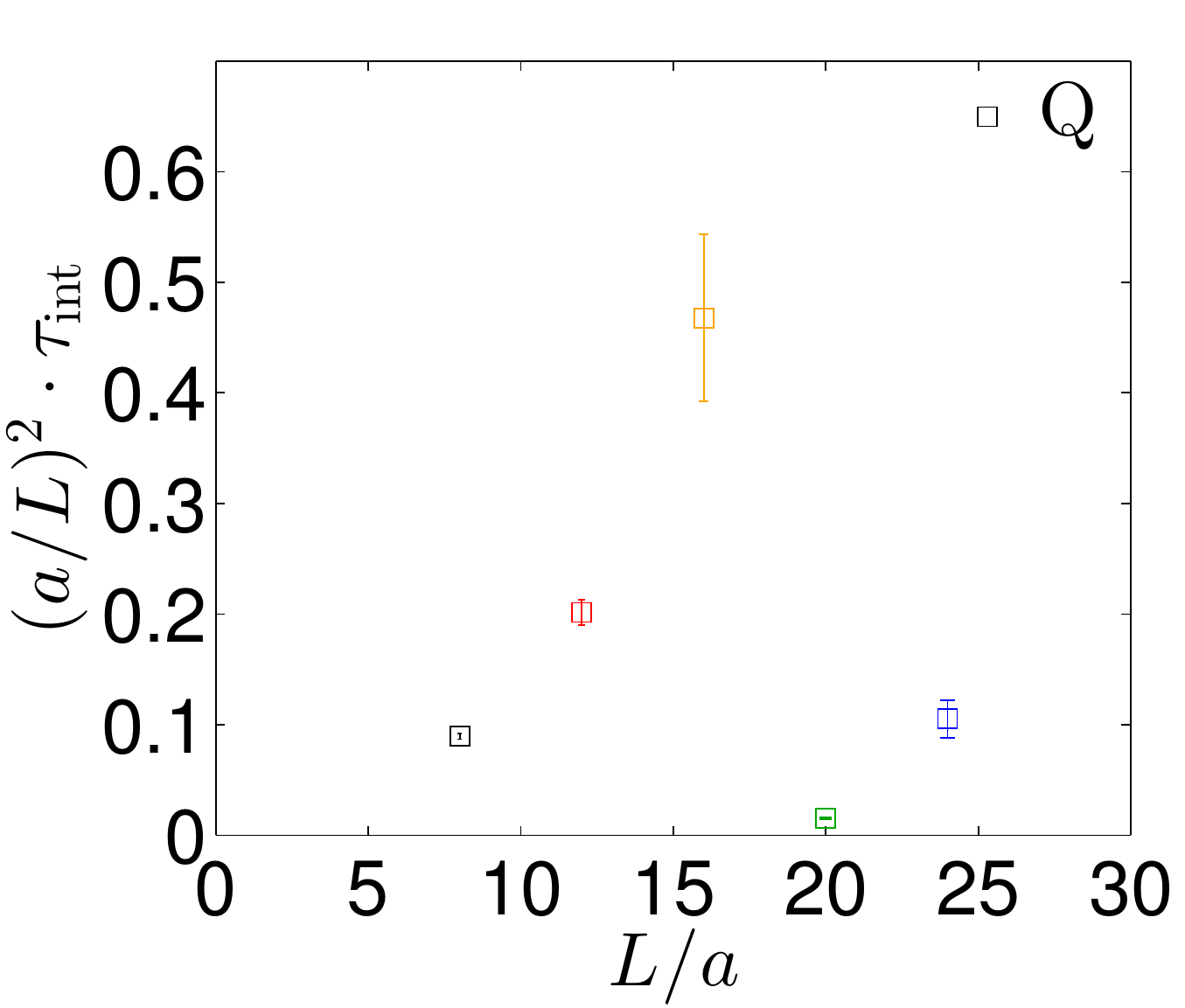}
\hdistance
\includegraphics[scale=\scaleB]{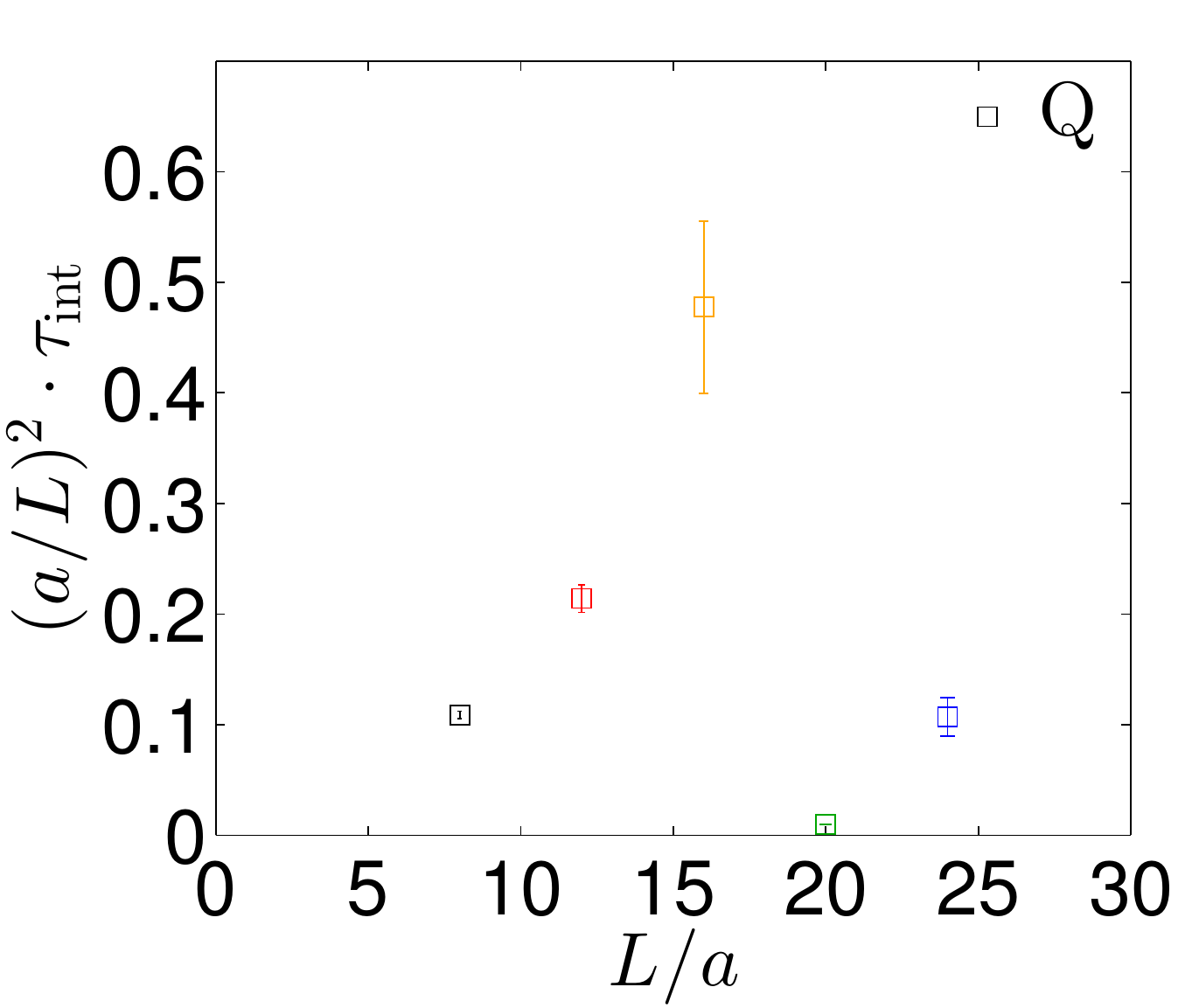} \\
\vdistance
\includegraphics[scale=\scaleB]{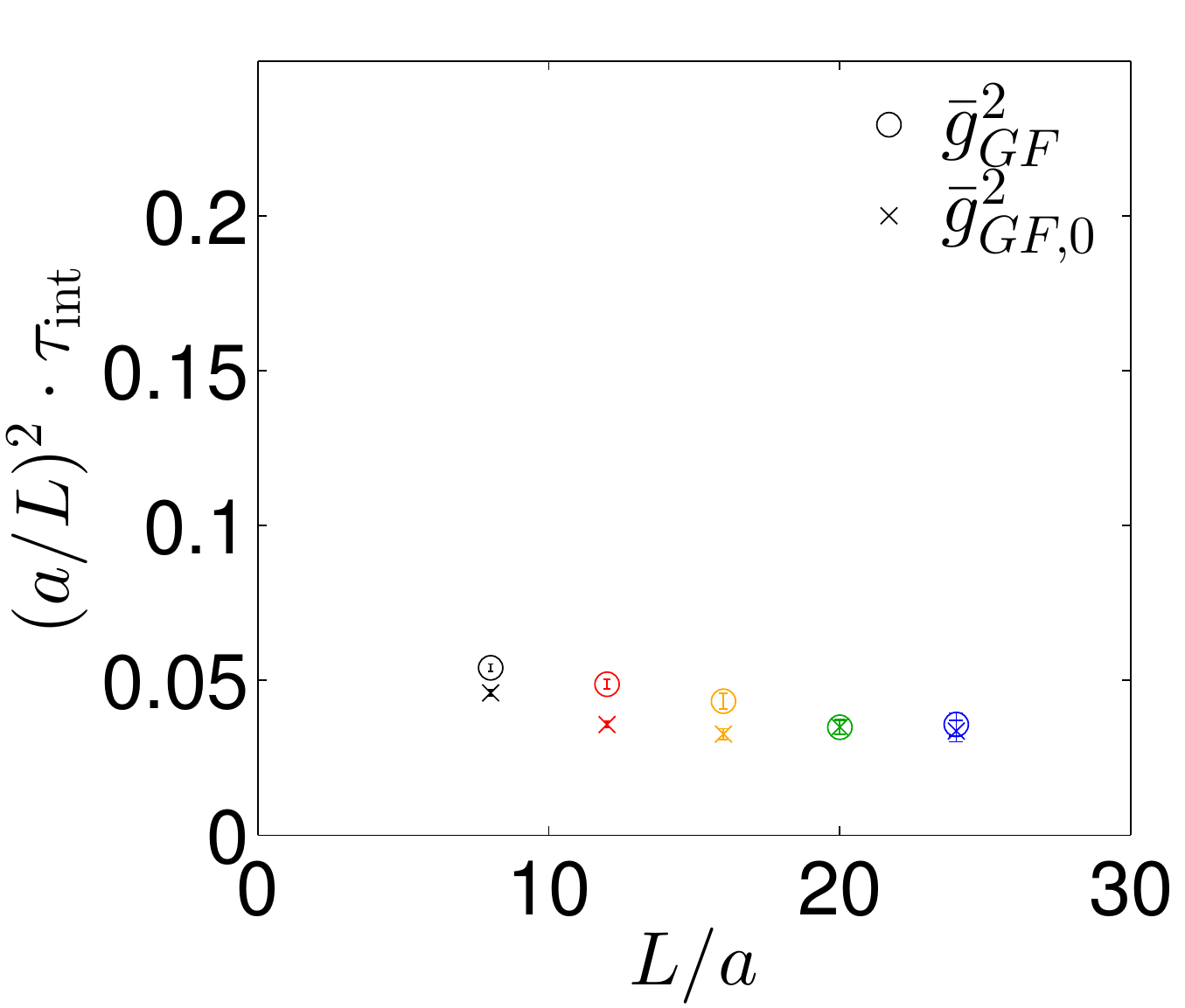}
\hdistance
\includegraphics[scale=\scaleB]{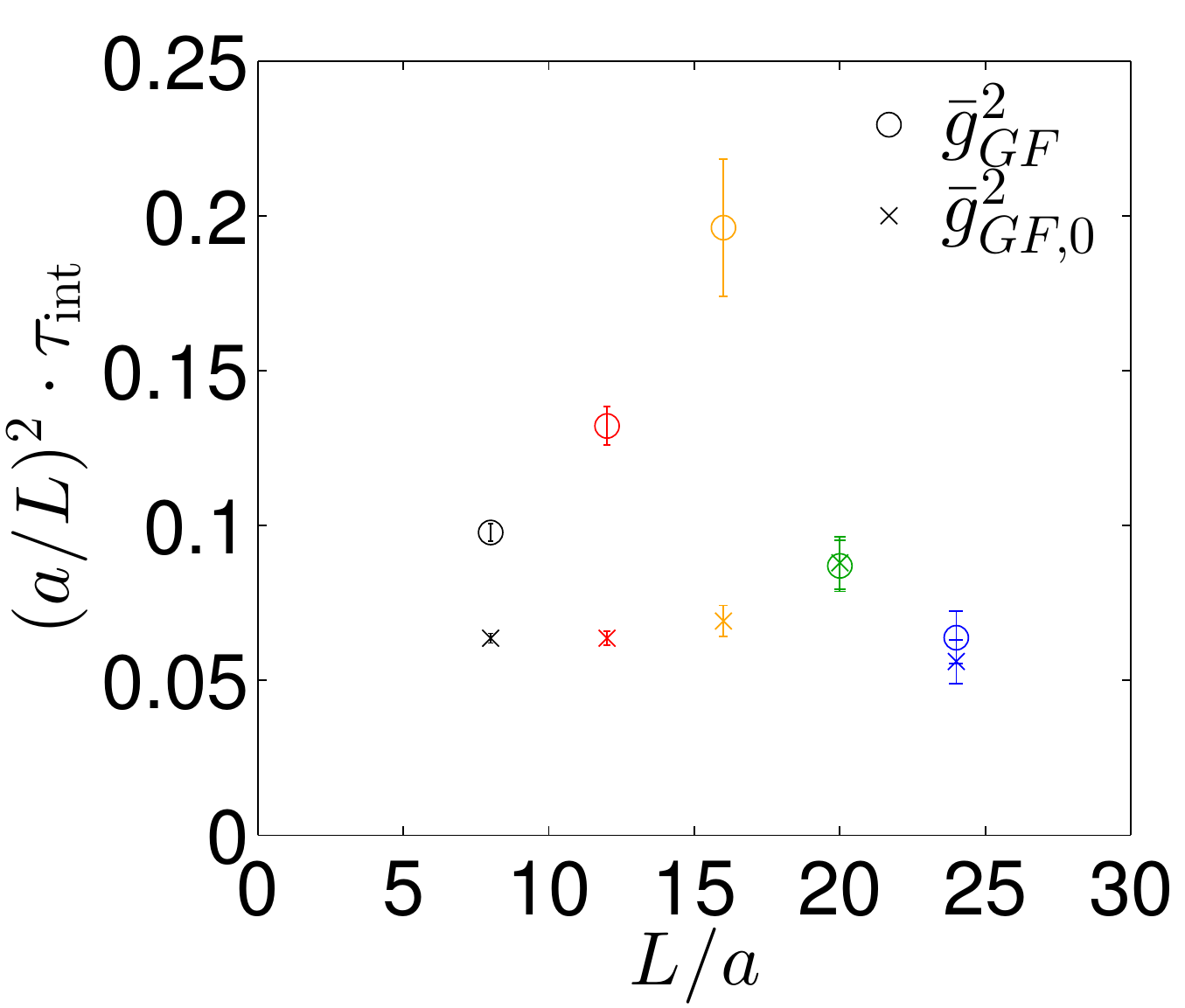}
\caption{Integrated autocorrelation time of the topological charge $Q$ (top) and the (modified) gradient flow coupling
(bottom), multiplied by $(a/L)^2$ and in units of 2 MDU.
         The left panel corresponds to $c=0.3$, the right one to $c=0.5$.
         }
\label{fig:autocorrelation_Q}
\end{figure}
For the largest lattices $L/a=20,24$, configurations from non-trivial sectors appear less often,
see 
Tab.~\ref{tab:Q_distribution}. However, the autocorrelations of $Q$ are obviously largely underestimated
(cf. Fig.~\ref{fig:autocorrelation_Q}), from which we infer that the current statistics is not sufficiently large
to sample the topological sectors correctly. 
Note that throughout this work, statistical errors were computed using the $\Gamma$-method \cite{Madras:1988ei, Wolff:2003sm}.

\subsection{Correlation of gradient flow coupling and topological charge \label{ss:correlation}}

\begin{figure}
\centering
\hspace{1.5mm}
\includegraphics[scale=\scaleB]{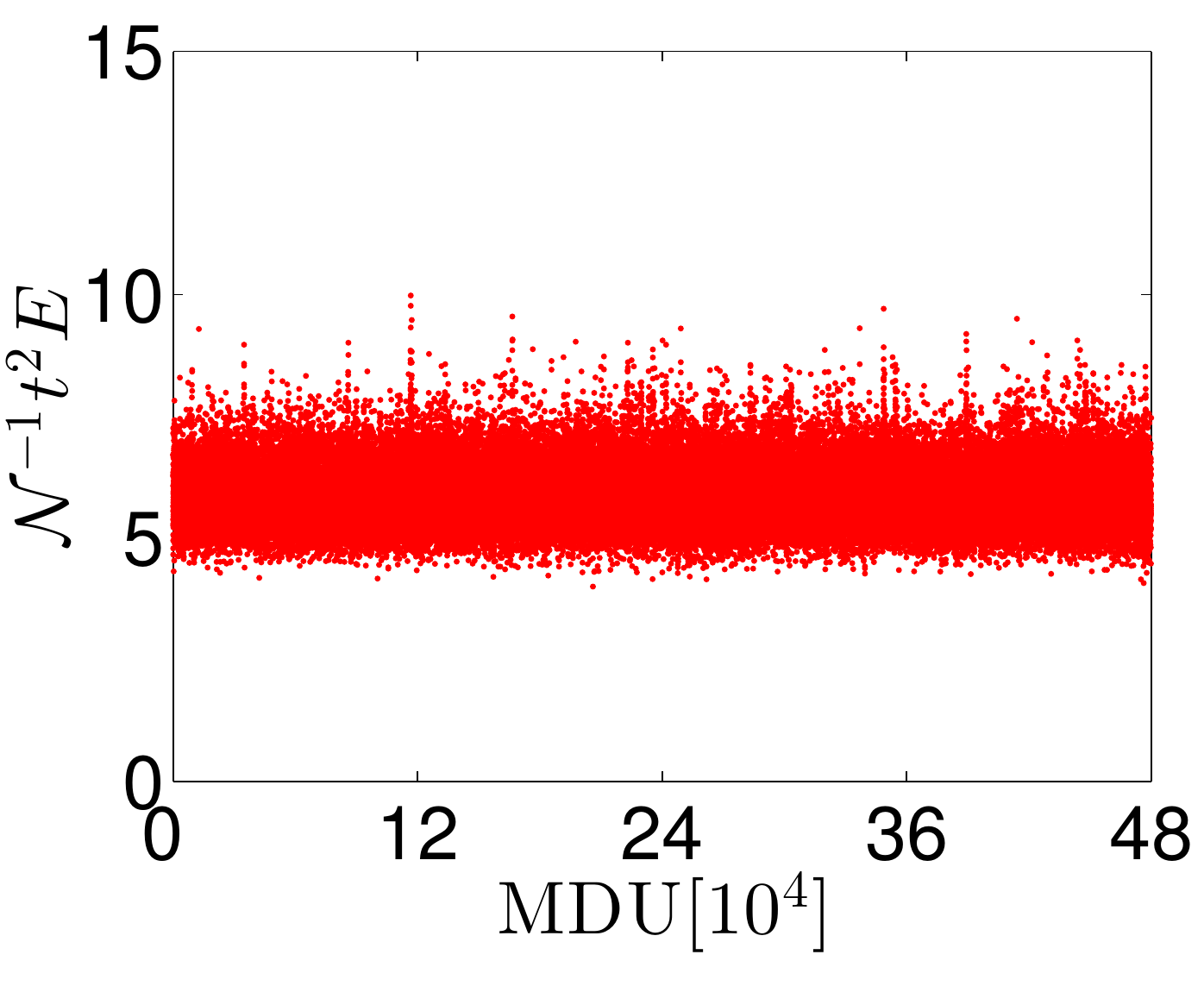}
\hdistance \hspace{-2.5mm}
\includegraphics[scale=\scaleB]{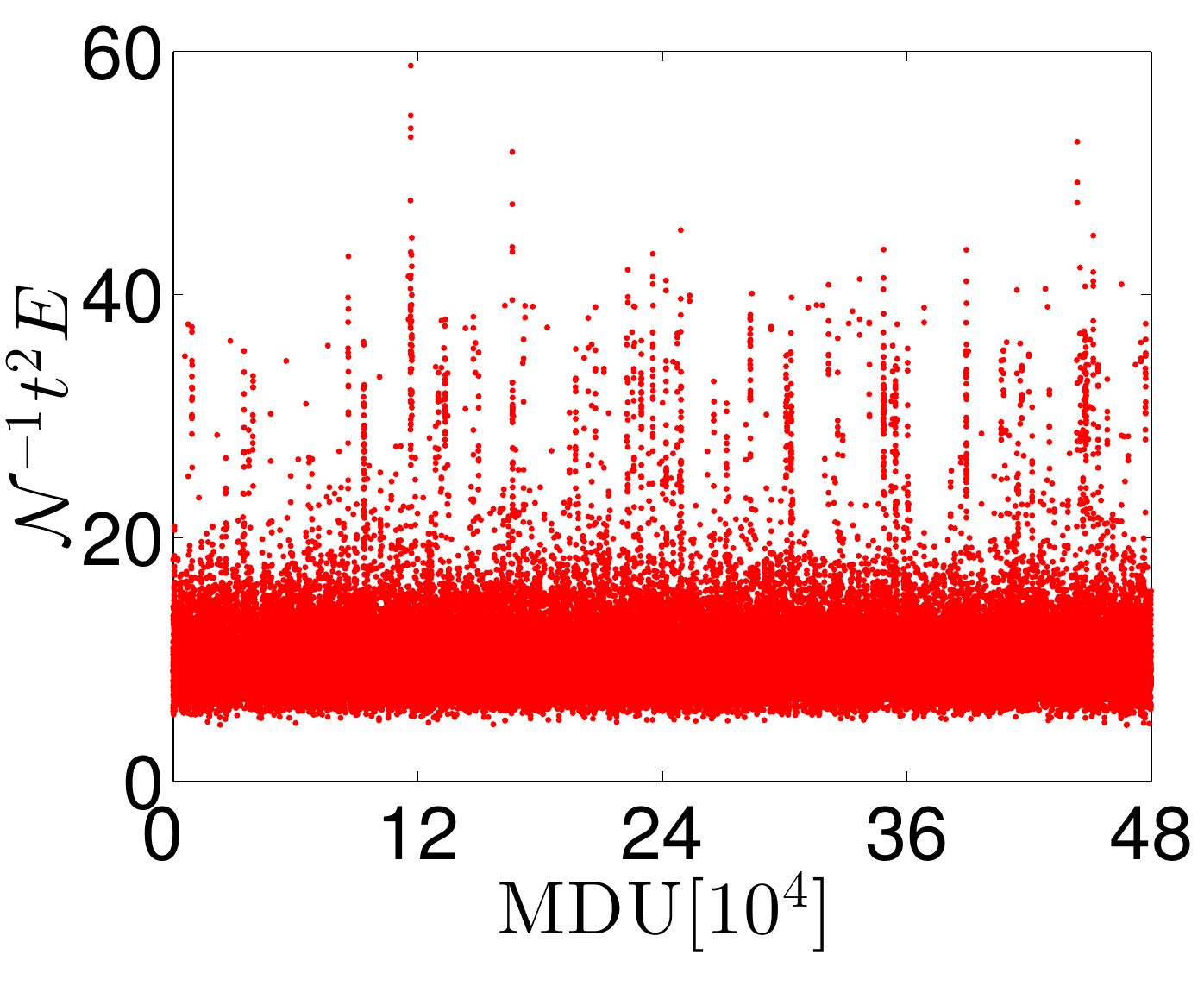} \\
\vdistance
\includegraphics[scale=\scaleB]{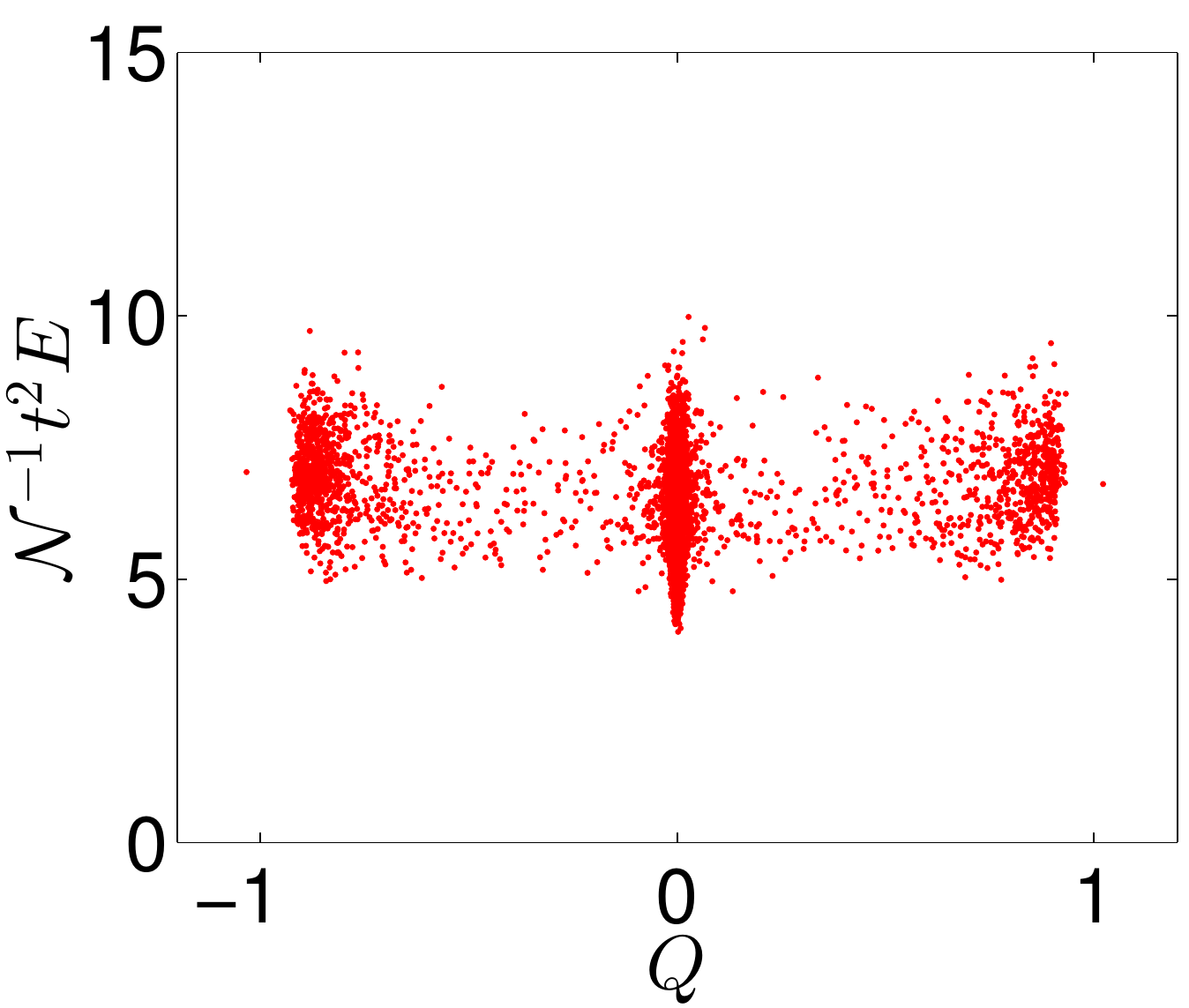}
\hdistance 
\includegraphics[scale=\scaleB]{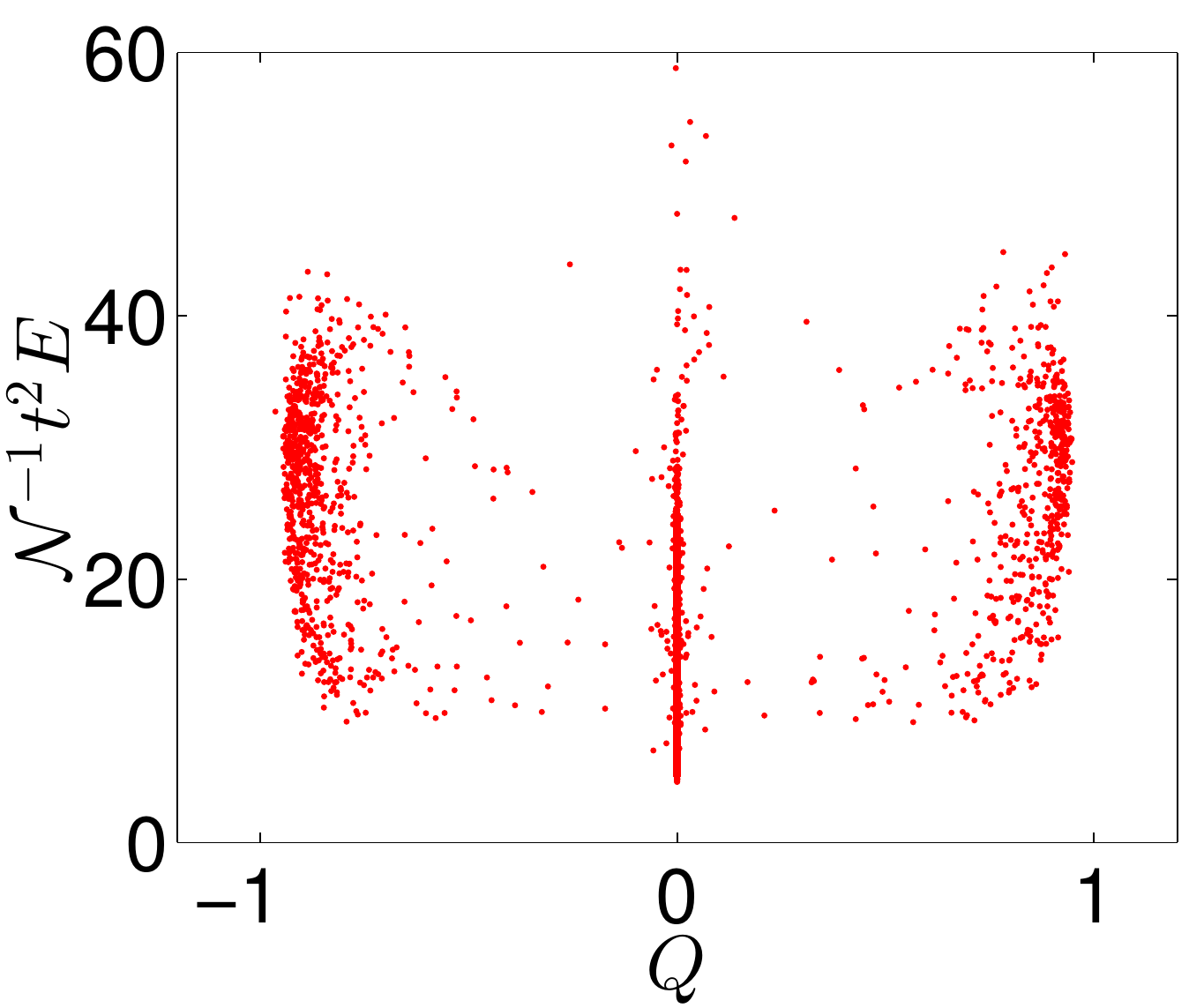} 
\caption{Top: Histories of $\bar g_{\rm GF}^2$.
         Bottom: Correlation of $\bar g_{\rm GF}^2$ and $Q$.
         All plots are shown exemplarily for $L/a=12$.
         The left panel corresponds to $c=0.3$, the right one to $c=0.5$.
         }
\label{fig:g_histories_12}
\end{figure}

The histories of the quantity $\mathcal N^{-1} t^2 E$, whose expectation value is the gradient flow
coupling $\bar g_{\rm GF}^2$, exhibit a certain amount of large values, see Fig.~\ref{fig:g_histories_12}.
This phenomenon is more pronounced the larger $c$ is chosen, and goes along with an increasing correlation
between the gradient flow coupling and the topological charge. 
In particular, the large values of $\mathcal N^{-1} t^2 E$ stem to a high
amount from configurations of non-vanishing topological charge.
In turn, the correct sampling of the topological sectors becomes a necessity in order
to obtain correct results. 
However, as we have seen in sec.~\ref{ss:Q}, this requirement is not fulfilled 
for the two largest lattices,
which means that the results for $\bar g_{\rm GF}^2$
on these lattices are biased. 

\subsection{The modified GF coupling}

\begin{figure}
\centering
\includegraphics[scale=\scaleB]{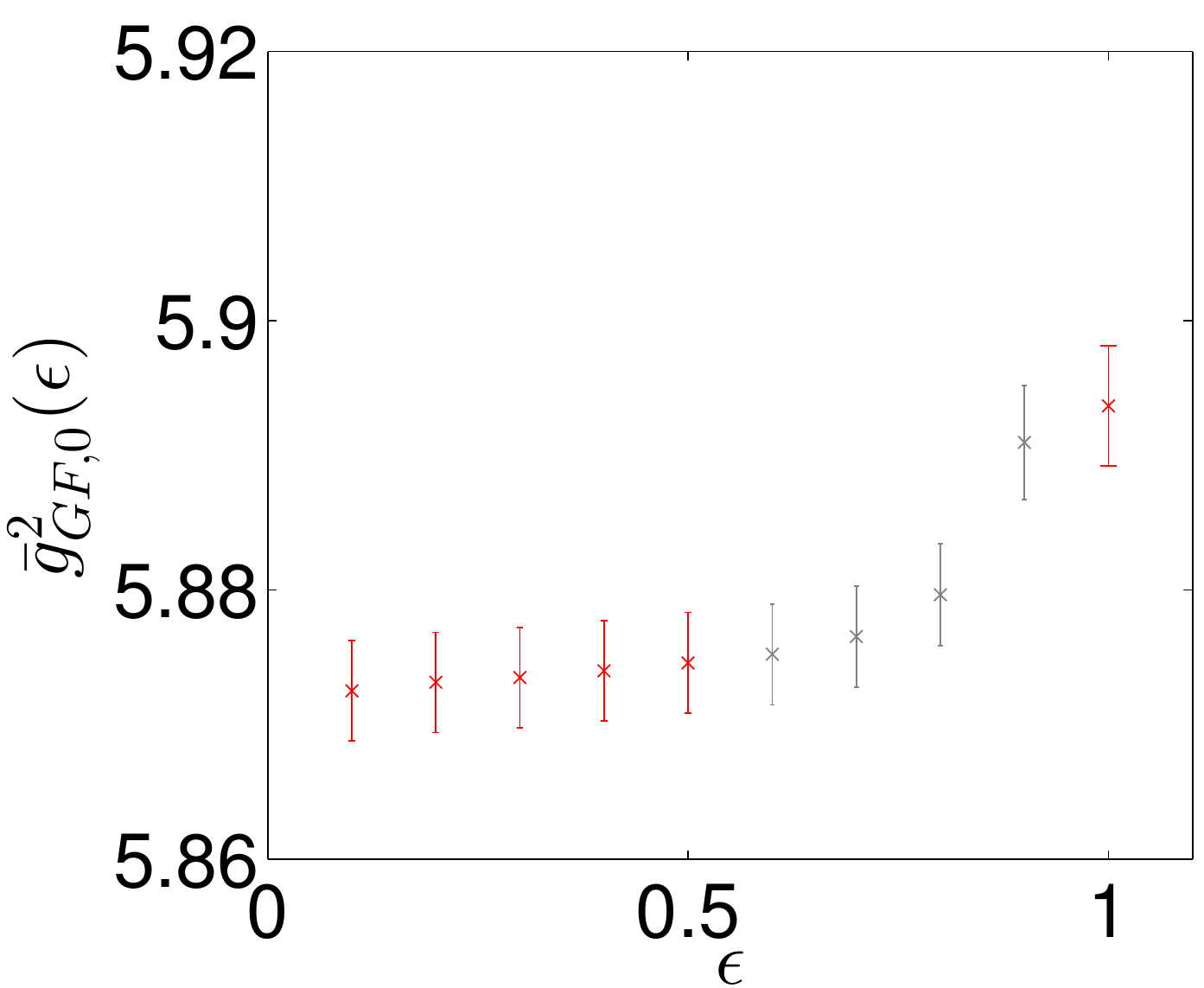}
\hdistance
\includegraphics[scale=\scaleB]{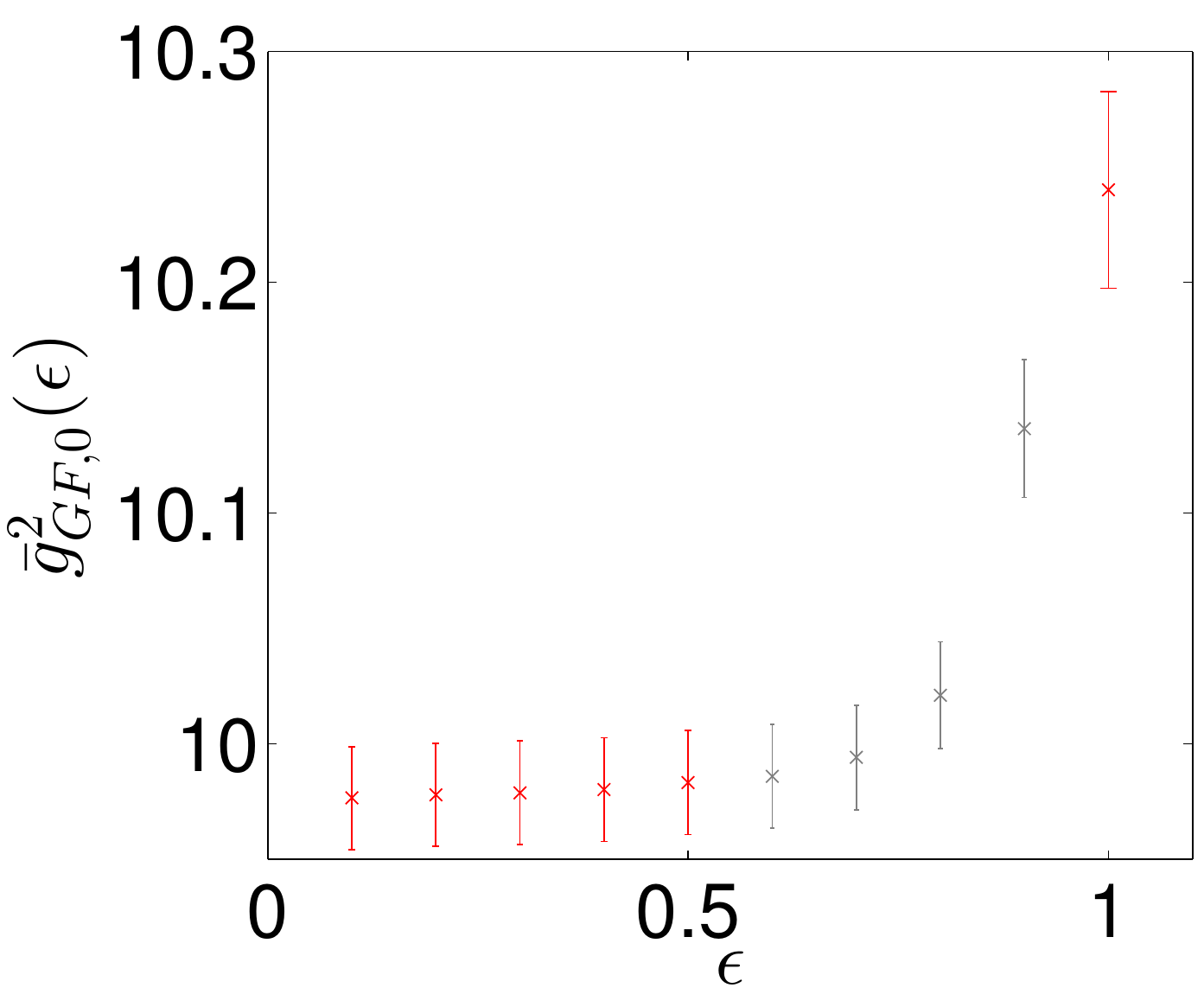}
\caption{Modified gradient flow coupling $\bar g_{\rm GF,0}^2$ for $L/a=12$ against the range $\epsilon$, where $\epsilon \leq 0.5$ can serve as
         definition of the trivial sector
         on the lattice. Results for larger $\epsilon$ take into account configurations which are considered
         non-trivial, but are nevertheless shown in gray for completeness. The point at $\epsilon = 1$ corresponds to
         the original definition $\bar g_{\rm GF}^2$ of the gradient flow coupling. Left: $c=0.3$. Right: $c=0.5$.}
\label{fig:gmod}
\end{figure}

In order to assess the impact of the non-trivial topological sectors and their insufficient sampling
on the determination of $\bar g_{\rm GF}^2$,
we consider a modified GF coupling, which has the same perturbative expansion but takes into account only gauge configurations from the trivial sector:
\begin{equation}
  \bar g_{\rm GF,0}^2 = \mathcal N^{-1} ~ t^2 ~\frac{\langle E(t) ~\delta_{Q,0} \rangle}{\langle \delta_{Q,0} \rangle}\bigg|_{t=c^2 L^2/8} \ . 
  \label{eq:gmod}
\end{equation}
On the lattice, where we have non-integer values of $Q$, all configurations with $|Q| \leq \epsilon$  
($\epsilon = 0.1, 0.2,$ \newline $\ldots, 0.5$) are considered to belong to the 
trivial sector, i.e., we replace
$
 \delta_{Q,0} \to \Theta(Q+\epsilon)\Theta(\epsilon-Q)
$
in (\ref{eq:gmod}).
The results for $L/a=12$ can be seen in Fig.~\ref{fig:gmod}.
The contributions from non-trivial sectors do make a difference, 
and the effect is
stronger for large $c$ due to the larger correlation discussed in sec.~\ref{ss:correlation}.
Moreover, we see that the particular choice of $\epsilon$
has no big influence on the modified gradient flow coupling. We use $\epsilon = 0.5$ in the following.
In Fig.~\ref{fig:autocorrelation_Q}, we compare the integrated autocorrelation time for the gradient flow
coupling in its original and modified form.
We find that the original coupling is affected by the bad sampling towards the continuum, whereas
the modified coupling does suffer less severely from critical slowing down and shows the expected {$\sim 1/a^2$} behavior.
In that sense, the modified gradient flow coupling can be considered to be safer.

\subsection{Results and continuum limit}

The full set of results for the two couplings $\bar g_{\rm GF}^2$ and $\bar g_{\rm GF,0}^2$ is listed in Tab.~\ref{tab:continuum_limit}.
On the coarser
lattices ($L/a=8,12,16$), the simulations show a
clear difference between the two definitions. 
This suggests that in the studied volume \mbox{($L\sim 0.8\,{\rm
    fm}$)}, topologically non-trivial configurations play a role in
accurately determining the value of $\bar g_{\rm GF}^2$.
On the two finer lattices ($L/a=20,24$), we do not observe a difference
due to the critical slowing down that affects the determination of the original coupling.

\begin{figure}
\centering
\includegraphics[scale=\scaleB]{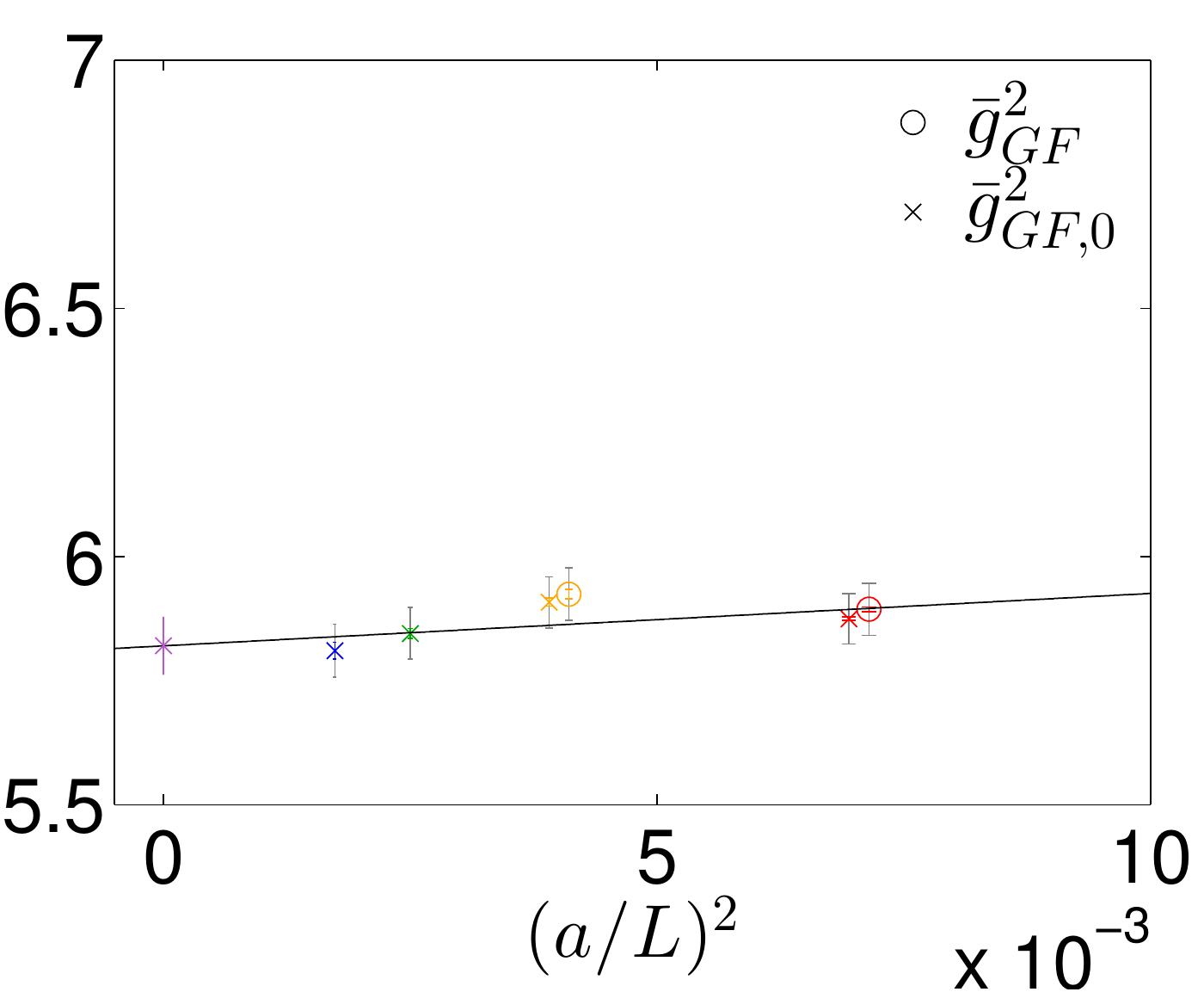}
\hdistance
\includegraphics[scale=\scaleB]{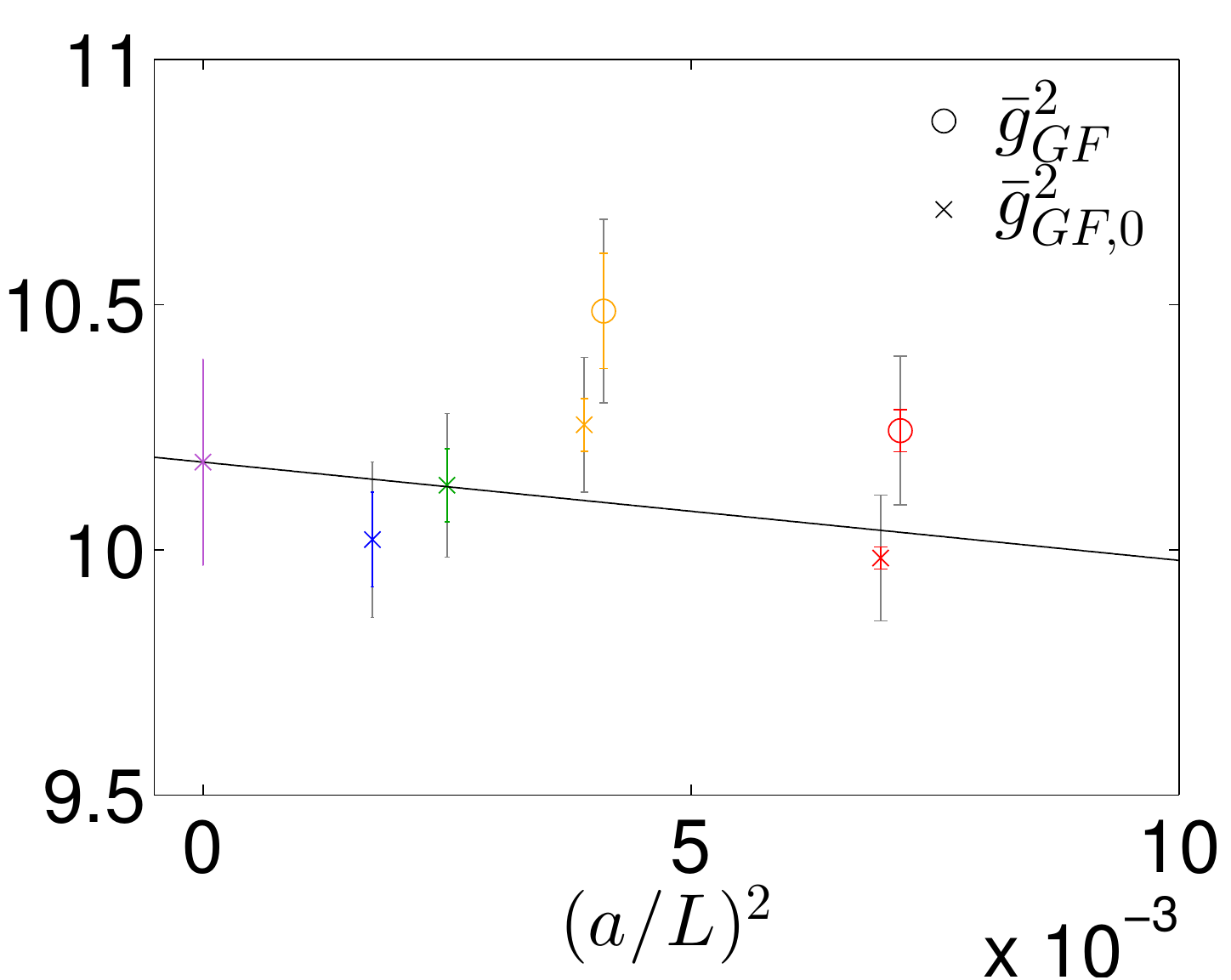}
\caption{
         Continuum extrapolation of the modified gradient flow coupling (crosses).
         The unbiased data for the gradient flow coupling is also shown (open circles), the points
         being slightly shifted to the right for convenience.
         Statistical errors are displayed in color, errors from the line of constant physics in gray.
         Left: $c=0.3$. Right: $c=0.5$.}
\label{fig:continuum_limit}
\end{figure}

Since the results for $\bar g_{\rm GF}^2$  on the
finer lattices ($L/a=20, 24$) are biased, we conduct the continuum extrapolation 
only for $\bar g_{\rm GF,0}^2$. To compare data
of different lattice spacings we have to take into account an
additional error  
being introduced by the way the physical volume was fixed,
Eq.~(\ref{eq:scalesetting})\footnote{The error on $a/r_0(\beta)$ 
  depends on $\beta$, but for simplicity we prop:qagate its
  maximum value of $1\%$ globally
  \cite{necco01_latticescaleNf0}.}. This uncertainty turns out to be  
larger than the statistical errors (see Tab.~\ref{tab:continuum_limit}).
We find that the data with $L/a>8$ is well described by a fit linear
in $(a/L)^2$, see 
Fig.~\ref{fig:continuum_limit}.

\begin{table}[h]
\centering
\small
\begin{tabular}{c|c|c|c|c}  
              & \multicolumn{2}{c|}{$c=0.3$} & \multicolumn{2}{c}{$c=0.5$}         \\ \cline{2-5}
   $L/a$      & $\bar g^2_{\rm GF}$          & $\bar g^2_{\rm GF,0}$   & $\bar g^2_{\rm GF}$   \vphantom{\"Ag} & $\bar g^2_{\rm GF,0}$ \\ \hline
  {$8$}       & {$5.647(\w{1}3)$}{(52)}      & {$5.631(\w{1}3)$}{(51)} & {$\w{1}9.313(\w{1}21)$}{(146)}  & {$\w{1}9.140(14)$}{(126)} \\
  {$12$}      & {$5.894(\w{1}5)$}{(52)}      & {$5.875(\w{1}4)$}{(51)} & {$10.243(\w{1}43)$}{(146)}      & {$\w{1}9.983(23)$}{(126)} \\
  {$16$}      & {$5.924(10)$}{(52)}          & {$5.908(\w{1}8)$}{(51)} & {$10.487(117)$}{(146)}          & {$10.255(54)$}{(126)}     \\
  {$20$}      & {*${5.845}({10})$}{(52)}     & {$5.845(10)$}{(51)}     & {*${10.135}(\w{1}{74})$}{(146)} & {$10.132(74)$}{(126)}     \\
  {$24$}      & {*${5.818}({28})$}{(52)}     & {$5.810(17)$}{(51)}     & {*${10.128}({123})$}{(146)}     & {$10.021(96)$}{(126)}     \\ \hline
  {$\infty$}  & {*$5.818(62)$}               & {$5.820(58)$}           & {*$10.156(258)$}                & {$10.179(210)$}  
\end{tabular}
\caption{Results for the (modified) gradient flow coupling and its continuum extrapolation. 
         The first error is statistical, the second one stems from fixing the physical volume.
         Biased values are denoted with an asterisk.}
\label{tab:continuum_limit}
\end{table}

\section{Summary \& Conclusions}

We investigated the gradient flow coupling in pure SU(3) Yang-Mills theory in a volume
of \mbox{$L \sim 0.8~{\rm fm}$}. 
We find a significant correlation between the coupling and the
topological charge, which 
increases in strength with the smoothing fraction $c$. 
Simulations suffering from a bad sampling of topological sectors and
critical slowing down would lead to a biased determination of {$\bar
  g^2_{\rm GF}$} in the continuum. We propose an alternative
definition for the coupling ($\bar g^2_{\rm GF,0}$), which takes into
account only the trivial topological sector, and show that its
determination is not affected by the bad topology sampling in the studied
volume. We think that the alternative definition might be advantageous
for the determination of the running coupling in intermediate
volumes.


\begin{thebibliography}{99}
  \bibitem{luescher10_flow} M. L\"uscher. JHEP 1008 (2010) 071,
    \texttt{[arXiv:1006.4518]};\\
    M.~L{\"u}scher and P.~Weisz. JHEP 1102 (2011) 051, [{{\tt
        arXiv:1101.0963}}] 
  \bibitem{Fodor:2012td} Z. Fodor et al. JHEP 1211 (2012) 007,
    \texttt{[arXiv:1208.1051]} 
  \bibitem{fritzsch13_flowNf2} P. Fritzsch, A. Ramos. JHEP 1310 (2013)
    008, \texttt{[arXiv:1301.4388]}  
  \bibitem{Schaefer:2010hu} S.~Schaefer {\it et al.} Nucl.\ Phys.\ B {\bf 845}, 93 (2011) [{{\tt arXiv:hep-lat/1009.5228}}]
  \bibitem{necco01_latticescaleNf0} S. Necco, R. Sommer. Nucl. Phys. B622 (2002) 328-346, \texttt{[arXiv:hep-lat/0108008]}
  \bibitem{luescher_openQCD} M. L\"uscher, S. Schaefer. JHEP 1107 (2011) 036, \texttt{[arXiv:1105.4749]}
  \bibitem{Bode:1999sm} A. Bode, P. Weisz, U. Wolff. Nucl. Phys. B576 (2000) 517-539, \texttt{[arXiv:hep-lat/9911018]}
  \bibitem{Madras:1988ei} N. Madras, A. D. Sokal. J.Statist.Phys. 50 (1988) 109-186 
  \bibitem{Wolff:2003sm} U. Wolff. Comput.Phys.Commun. 156 (2004) 143-153, \texttt{[arXiv:hep-lat/0306017]}
  \end{thebibliography}
\end{document}